\documentclass[final,authoryear,5p,times,twocolumn]{elsarticle}
\usepackage{amssymb}
\biboptions{square,comma,numbers,sort&compress}

\journal{Physica Scripta}

\usepackage{epsfig}
\usepackage{psfrag}
\newcommand{\mypsfrag}[2]{\psfrag{#1}{\footnotesize{#2}}}
\usepackage{amsfonts}          % |
\usepackage{amsmath}           % |
\usepackage[a4paper,ps2pdf,colorlinks=true,breaklinks=true]{hyperref}
\usepackage{ulem}
\usepackage{xspace}
\renewcommand{\vec}[1]{\mbox{\boldmath $ #1$}}

\newcommand{\x}{\cdot}
\providecommand{\Mdip}{\ensuremath{{M}_\text{dip}}\xspace}

\newcommand{\sgn}{\text{sgn}}
\renewcommand{\Re}{\text{Re}}
\renewcommand{\Im}{\text{Im}}

\usepackage{color}

\newcommand{\revb}[1]{}
\newcommand{\hide}[1]{}

\allowdisplaybreaks

\newlength{\shortwidth}
\newlength{\textwidtha}
\begin{document}

\begin{frontmatter}



\title{Solar cycle properties described by simple convection-driven dynamos}


\author{
\textbf{Radostin D~Simitev}$^{1,3}$
and 
\textbf{Friedrich H Busse}$^{2,3}$ 
}
\address{
$^1$
School of Mathematics and Statistics, University of Glasgow --
  Glasgow G12 8QW, UK, EU\\
$^2$
Institute of Physics,  University of Bayreuth -- Bayreuth
  D-95440, Germany, EU\\
$^3$ 
NORDITA, AlbaNova University Center -- Stockholm SE-10691,
  Sweden, EU\\[2mm]
E-mail: \href{mailto:busse@uni-bayreuth.de}{busse@uni-bayreuth.de}
}

\begin{abstract}
Simple models of magnetic field generation by convection in rotating
spherical shells exhibit properties resembling those observed on the
sun. The {assumption of the Boussinesq approximation made in these
models} prevents a  realistic description of the solar cycle, but
through a physically motivated change in the boundary condition for the differential
rotation the propagation of dynamo waves towards higher latitudes can
be reversed at least at low latitudes.  
\end{abstract}


\end{frontmatter}
%\linenumbers

\section{Introduction}

A well known difficulty in modeling the solar dynamo is the fact that
the dynamo waves which describe the nearly time periodic dynamics of
the magnetic field tend to propagate from lower to higher latitudes
instead in the opposite sense as observed on the sun. This effect is
well known in mean field models of the solar cycle, see for example
Stix (1976), but it is also observed in numerical solutions {of
convection-driven} dynamos in rotating spherical fluid shells which are
supposed to model processes in the solar convection zone. This
shortcoming and others are caused by the inadequate representation of
the compressibility of the solar atmosphere. Even with the huge power
of modern supercomputers it is not yet possible to resolve adequately
the dynamics of convection in the presence of the large density
variation between bottom and top of the solar convection zone and to
model appropriately the {dependence of density on pressure.}

{In their early} direct numerical models of solar convection and 
magnetic field generation, Gilman and Miller (1981) assumed the
Boussinesq approximation in which the fluid is regarded as
incompressible except in connection with the gravity term where the
temperature dependence of the density is taken into account. This
approximation eliminates the need for a separate equation of state and
leads to a system of equations describing long period processes while
the short period acoustic modes no longer enter the analysis. The same
effect is obtained in the anelastic approximation in which the
horizontally averaged density variation is taken into account, but the
fluctuating component of the density is still only represented in the
gravity term. For applications of the anelastic approximation in
models of the solar convection zone see Gilman and Glatzmaier (1981)
and Elliot et al. (2000). 

Convection in rotating spherical fluid shells heated from below is
always associated with a differential rotation generated by the
Reynolds stresses of convection. An analytical model demonstrating the
preference of banana shaped convection cells girdling the equator and
the associated solar like differential rotation was presented by Busse
(1970, 1973). While the analytical solution for stress-free boundaries
exhibits a depth independent differential rotation, a differential
rotation decreasing with depth is always found in fully nonlinear
numerical models. This property together with the fact that the
differential rotation reaches  its maximum at the equator is
responsible for the propagation of the dynamo waves towards higher
latitudes (Yoshimura, 1975).  
%fig. 1
\begin{figure*}
\psfrag{Omg}{$\Omega$}
\begin{center}
\epsfig{file=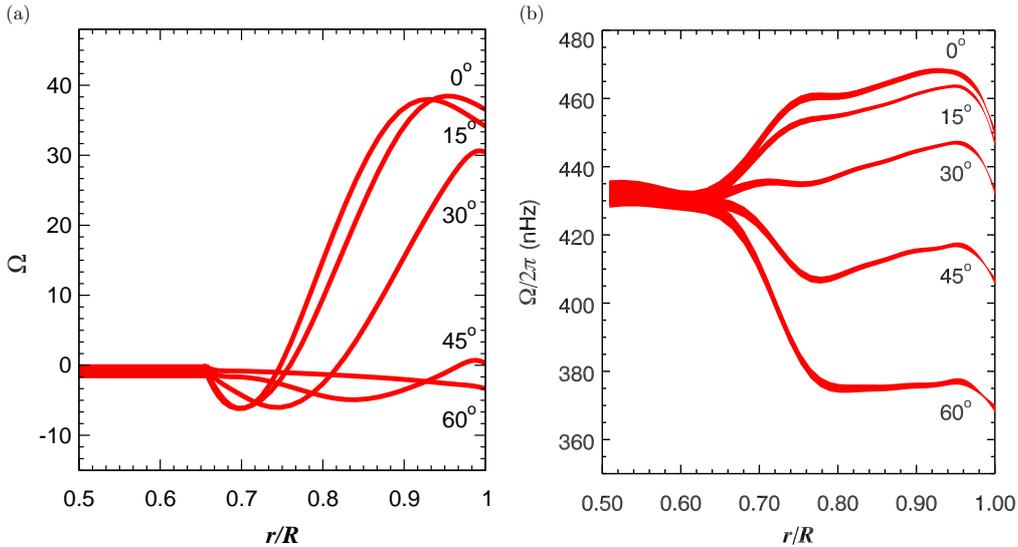,width=1.5\columnwidth,clip=}
\end{center}
\caption[]{{(color online)} Differential rotation of the solar convection zone
  according to helioseismology (right, image courtesy NSF's National
  Solar Observatory) and differential rotation of the numerical model 
in the case $\eta=0.65$, $P=1$, $\tau=2$, $R=120000$, $P_m=4$,
  $\beta=1.5$ and mixed velocity boundary conditions {(left)}.
}
\label{fig010}
\end{figure*}

The solar differential rotation also decreases with depth throughout
most of the convection zone with the exception of the tachocline
region near its bottom and a region near its surface as indicated in
figure \ref{fig010}(b). 
There is no general agreement about the origin of {the} upper 30Mm
deep layer in which the differential rotation increases with 
depth. Here we assume that it is caused by supergranular convection
that is characterized by a strong asymmetry between rising hot plasma
and descending cool plasma. This type of convection has been modeled
by hexagonal convection cells in the presence of rotation (Busse,
{2007}). As shown in this paper the asymmetry between rising and
descending flow in hexagonal convection cells does indeed generate a
differential rotation that increases with depth. We use this dynamical
property of convection as motivation to modify the usually assumed
stress-free boundary condition. Solely for the differential rotation
we apply the condition given in expression (A11) of the Appendix. The
resulting profiles shown in the example of figure \ref{fig010}(a) are
still not very solar like, but they show an increase with depth of the
differential rotation in the upper layer and  tend to generate a
more solar like behavior in the numerical simulations, {as
discussed below.}

\section{Simple {Convection-Driven} Dynamos}

Because convection driven dynamo solutions strongly depend on the
Prandtl number $P$ and on the magnetic Prandtl number $P_m$ for values
of the order unity and because even bistability has been found in this
parameter range (Simitev and Busse, 2009) we use a numerical model
with a minimum {number} of external parameters as outlined in the
Appendix. In this way it is possible to cover a relevant parameter
region in sufficient detail. In addition to the two Prandtl numbers
and the parameter $\beta$ introduced in the boundary condition (A11),
only the rotation parameter $\tau$ and the Rayleigh number $R$ must be
considered.   

In choosing parameter values for a numerical solar model the task is
most easily accomplished in the case of $\tau$ since only a value of
the (eddy-) viscosity must be chosen. Using a commonly accepted eddy
viscosity of the order $10^8 \text{m}^2/\text{s}$ (Gilman, 1983) we find
$\tau\approx2\cdot10^3$. It is more difficult to select an appropriate
range for the Rayleigh number. We shall consider values that exceed
the critical values for the onset of dynamo action by less than a
factor of two. Otherwise convection motions become too chaotic and the
structures of the dynamo solutions become less regular {and more
difficult to understand.}
%fig.2
\begin{figure*}[ht]
\begin{center}
\epsfig{file=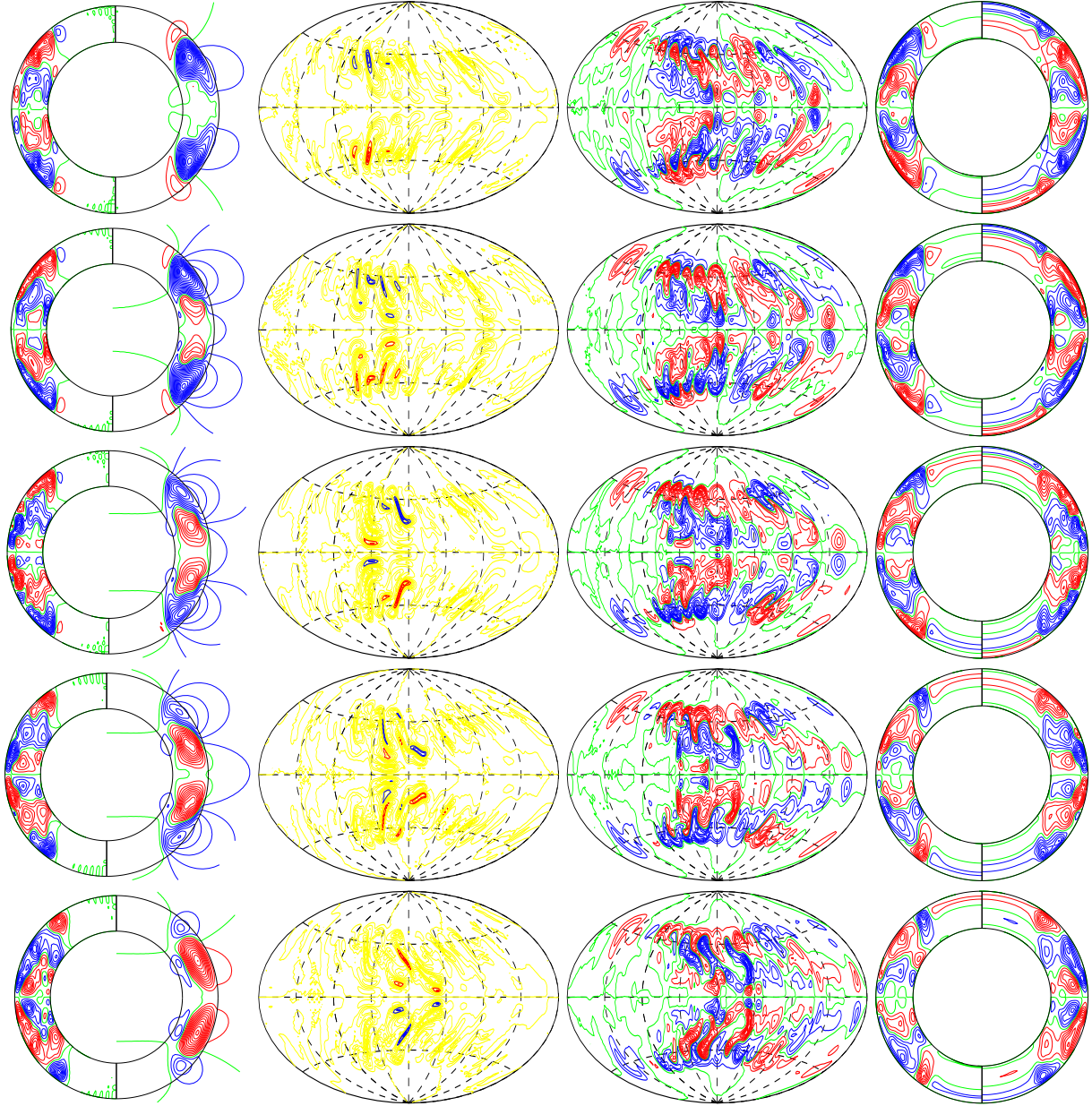,width=1.8\columnwidth,clip=}
\end{center}
\caption[]{{(color online)}Approximately half a period of a dynamo oscillation in the case 
 $\tau=2000$, $R=120000$, $P=1$, $P_m=4$, $\beta=1.5$. The first
 column shows meridional lines of constant $\overline{B_{\varphi}}$ on
 the left and poloidal fieldlines, $r \sin\theta
 \partial\overline{h}/\partial\theta$ on the right. The second column shows
 lines of constant {$\partial g/\partial\theta$} at $r=0.9$
 corresponding to -0.9, -0.8, -0.7, 0.7, 0.8, 0.9 of the maximum
 absolute value of {$\partial g/\partial\theta$},  and the third column
 shows lines of constant $B_r$ at $r=r_o$. The last column shows
 {$\Re(\partial g^{m=1}/\partial\theta)$} on the left and
 {$\Im(\partial g^{m=1}/\partial\theta)$} on the right. The five
 rows are separated equidistantly in time by $\Delta t =0.0224$.} 
\label{fig020}
\end{figure*}

In figure \ref{fig020} the properties of a convection-driven dynamo
are indicated by following the nearly periodic magnetic field through
half a magnetic cycle. The left column shows on the left side lines of
constant magnetic flux density $\overline{B}_{\varphi}$ where the bar
indicates the average over the azimuthal coordinate $\varphi$. On the
right side of the meridional cross-section lines of the axisymmertic
poloidal field are shown. The next column indicates the places where
the toroidal field reaches its maximum strength near the
surface. Since the toroidal field provides by far the major
contribution to the horizontal magnetic flux density, eruptions of
magnetic bipolar regions occur most likely  at those places in the
case of the sun.
{The modification of the stress-free boundary condition given by
expression (A11) causes the differential rotation to increase with depth in the
uppermost layer of the shell. This is turn causes the ``magnetic
bipolar regions'' shown in the second column of figure \ref{fig020} to
drift in a solar-like fashion from higher latitudes towards the
equator in agreement with solar observations.}
It is {also} remarkable that non-axisymmetric components
characterized by the azimuthal wavenumber $m=1$ of the magnetic field
dominate over the axisymmetric component as is evident from the third
column which shows the strength of the radial magnetic field at the
surface of the sphere. The superposition of the axis- and
non-axisymmetric components actually represent the property that the
dynamo process is nearly confined to one {meridional} hemisphere
while the other hemisphere is almost field free.    
%fig. 3
\begin{figure}[t]
\begin{center}
\epsfig{file=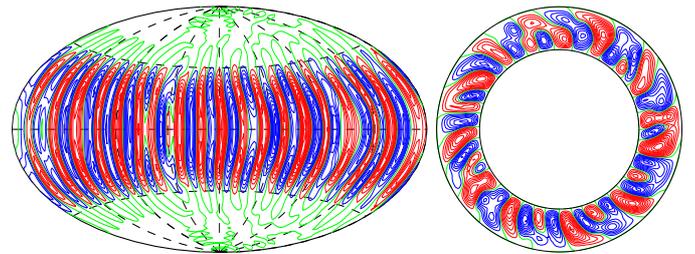,width=\columnwidth,clip=}
\end{center}
\caption[]{{(color online)}
Contour lines of the radial velocity $u_r$ at $r=r_i+0.5$ 
and the
poloidal streamlines $r \partial v/\partial\varphi$ in the equatorial plane
in the case
$P=1$, $\tau=2000$, $R=120000$, $P_m=4$, $\beta=1.5$.
}
\label{fig030}  
\end{figure}

The solar evidence for a non-axisymmetric magnetic field is provided
by the phenomenon of active longitudes (see Usoskin et al. (2007) and references therein).
 As in the case of those 
longitudes, the field shown in the third column of figure \ref{fig020} hardly
propagates throughout the time frame of the figure. The angular
velocity of the prograde propagation of the pattern is less than
$2\pi$ which is similar to the propagation of the convection
pattern. In contrast to {the} cyclical nature of the magnetic field the
convection pattern is nearly steady as indicated in figure
\ref{fig030}. Individual convection columns may sometimes breakup in
that the outer part tends to propagate a bit faster than the inner
part which essentially remains steady in the rotating frame.  

%fig. 4
\begin{figure*}
\begin{center}
\epsfig{file=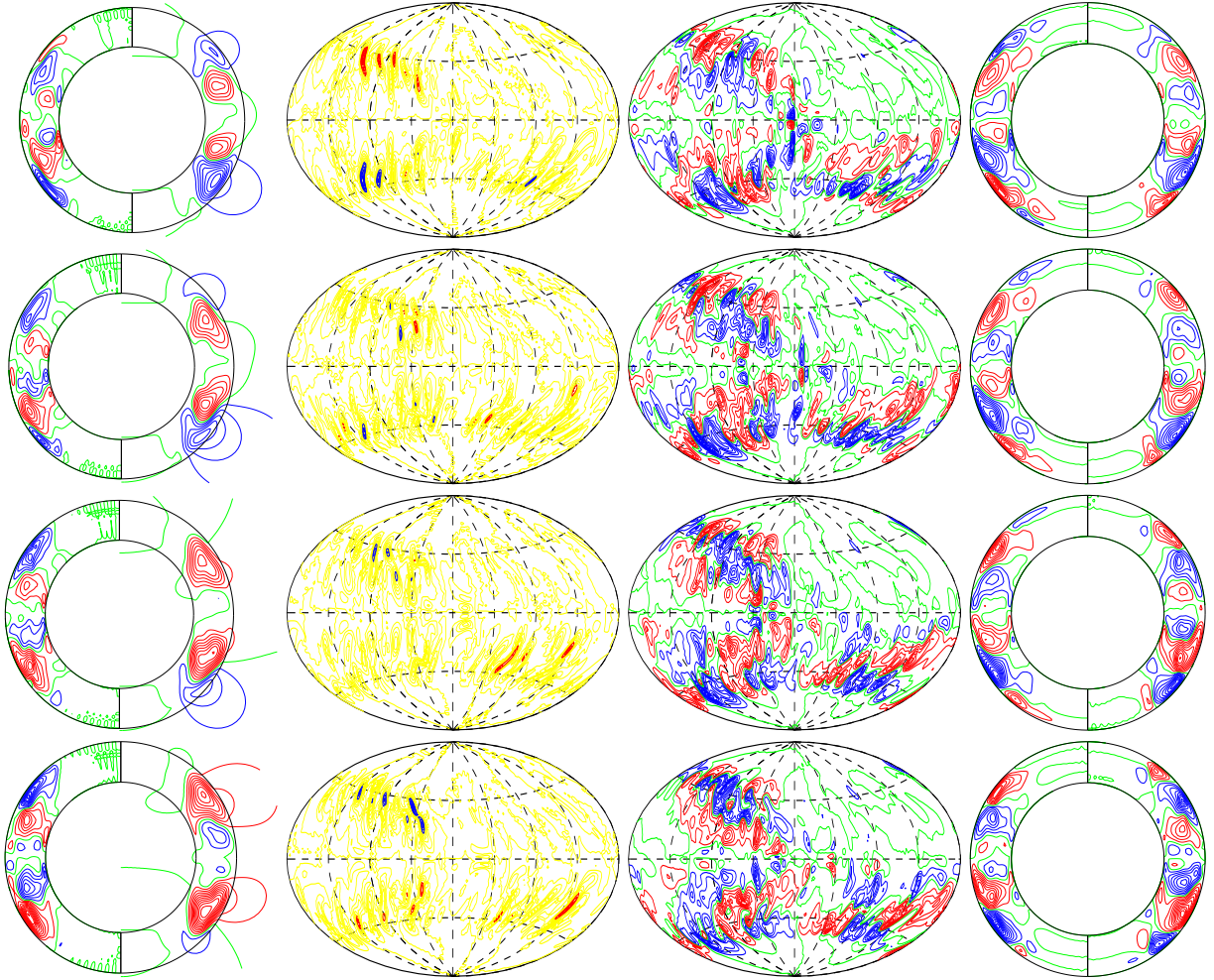,width=1.8\columnwidth,clip=}
\end{center}
\caption[]{{(color online)}Nearly half a period of a dynamo oscillation in the case 
 $\tau=2000$, $R=140000$, $P=1$, $P_m=3.5$, $\beta=1.0$. The first
 column shows meridional lines of constant $\overline{B_{\varphi}}$ on
 the left and poloidal fieldlines, $r \sin\theta
 \partial\overline{h}/\partial\theta$ on the right. The second column shows
 contour lines of $|B_{horizontal}|\sgn(B_{\varphi})$ at $r=0.9$
 corresponding to -0.9, -0.8, -0.7, 0.7, 0.8, 0.9 of the maximum value
 of $|B_{horizontal}|$,  and the third column shows lines of constant
 $B_r$ at $r=r_o$. The last column shows
 $\Re(\partial g^{m=2}/\partial \theta)$ on the left and $\Im(\partial
 g^{m=2}/\partial\theta)$ on the right.  The four rows are separated 
 equidistantly in time by $\Delta t =0.0168$.} 
\label{fig040}
\end{figure*}

\begin{figure}[t]
\mypsfrag{M}{\Mdip}
\mypsfrag{E}{$E$}
\mypsfrag{t}{$t$}
\epsfig{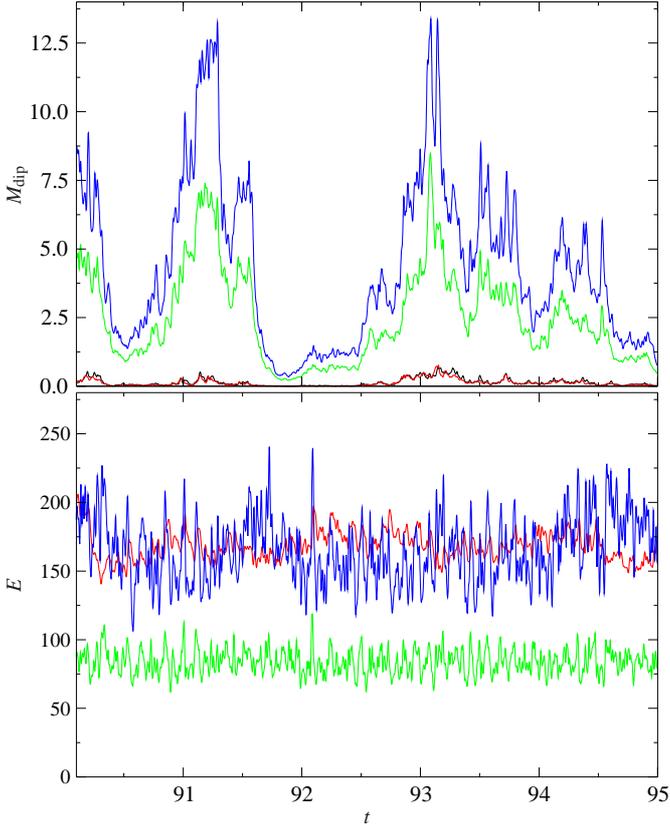}
\caption[]{{(color online)}
Time series of magnetic dipolar energy densities and kinetic energy
densities in the upper and lower panel respectively. 
The component $\overline{X}_p$ is shown by solid black 
line, while $\overline{X}_t$, $\check{X}_p$, and $\check{X}_t$
are shown by red, green and blue lines, respectively. $X$ stands
for either $\Mdip$ or $E$. The parameter values are $P=1$, $\tau=2000$, $R=140000$, $P_m=3.5$, $\beta=1$.
}
\label{fig050}  
\end{figure}

In figure \ref{fig040} half a magnetic cycle is presented for a higher
Rayleigh number than in the case of figure \ref{fig020}. As is evident
from the 3 column of figure \ref{fig040}  the structure of the magnetic field is
not antisymmetric with respect to the equatorial plane. In the
southern hemisphere the {$m=2$} --component of the magnetic field dominates
in this particular cycle. At other times patterns that are more
antisymmetric and similar to that shown in figure \ref{fig020} are
found. The amplitude of the magnetic field varies strongly from cycle
to cycle as shown in figure \ref{fig050}  while amplitude of convection exhibits
rather small variations.     

%fig. 5

%fig.6
\begin{figure}
\begin{center}
\epsfig{file=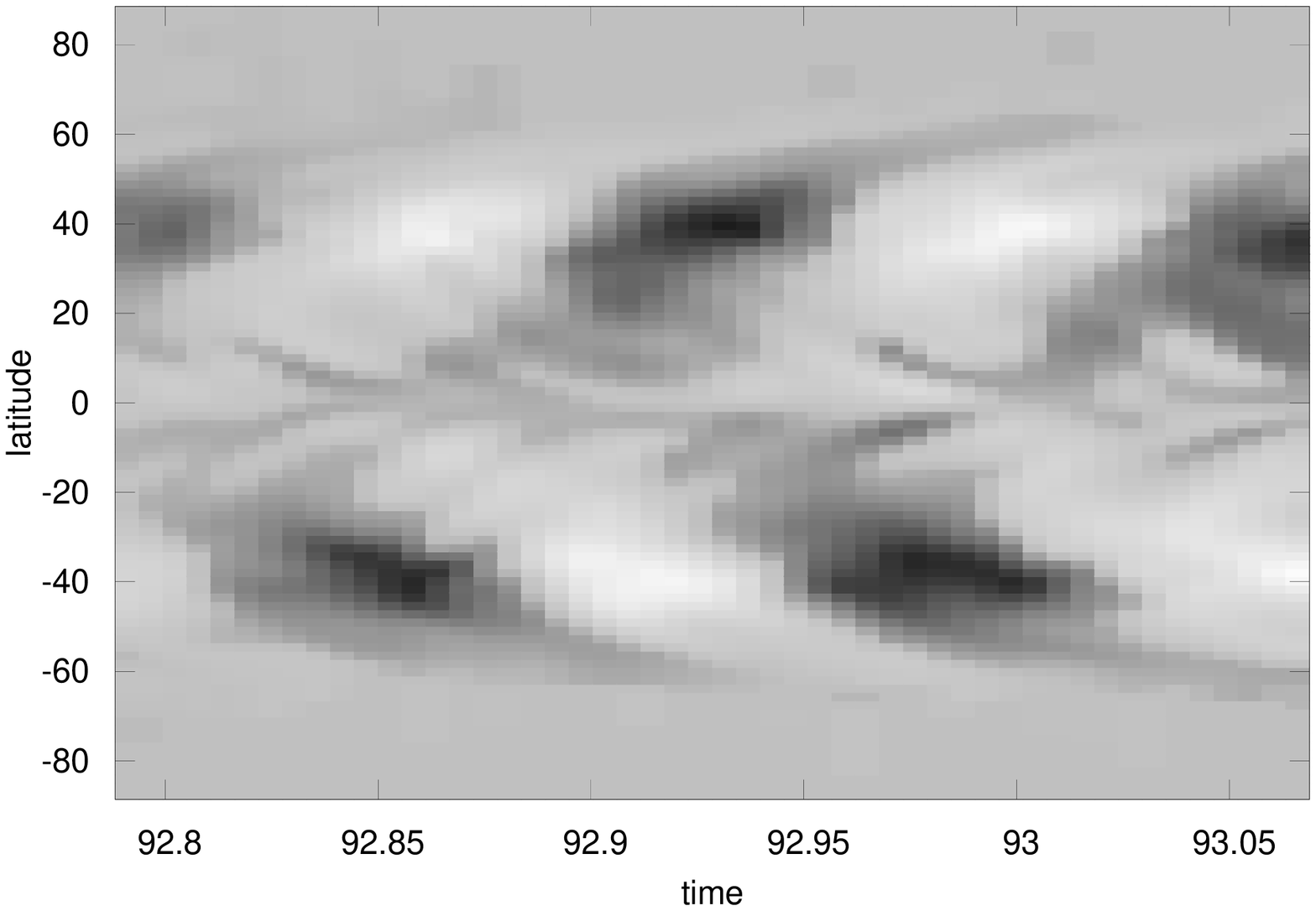,height=\columnwidth,angle=-90,clip=}\\
\epsfig{file=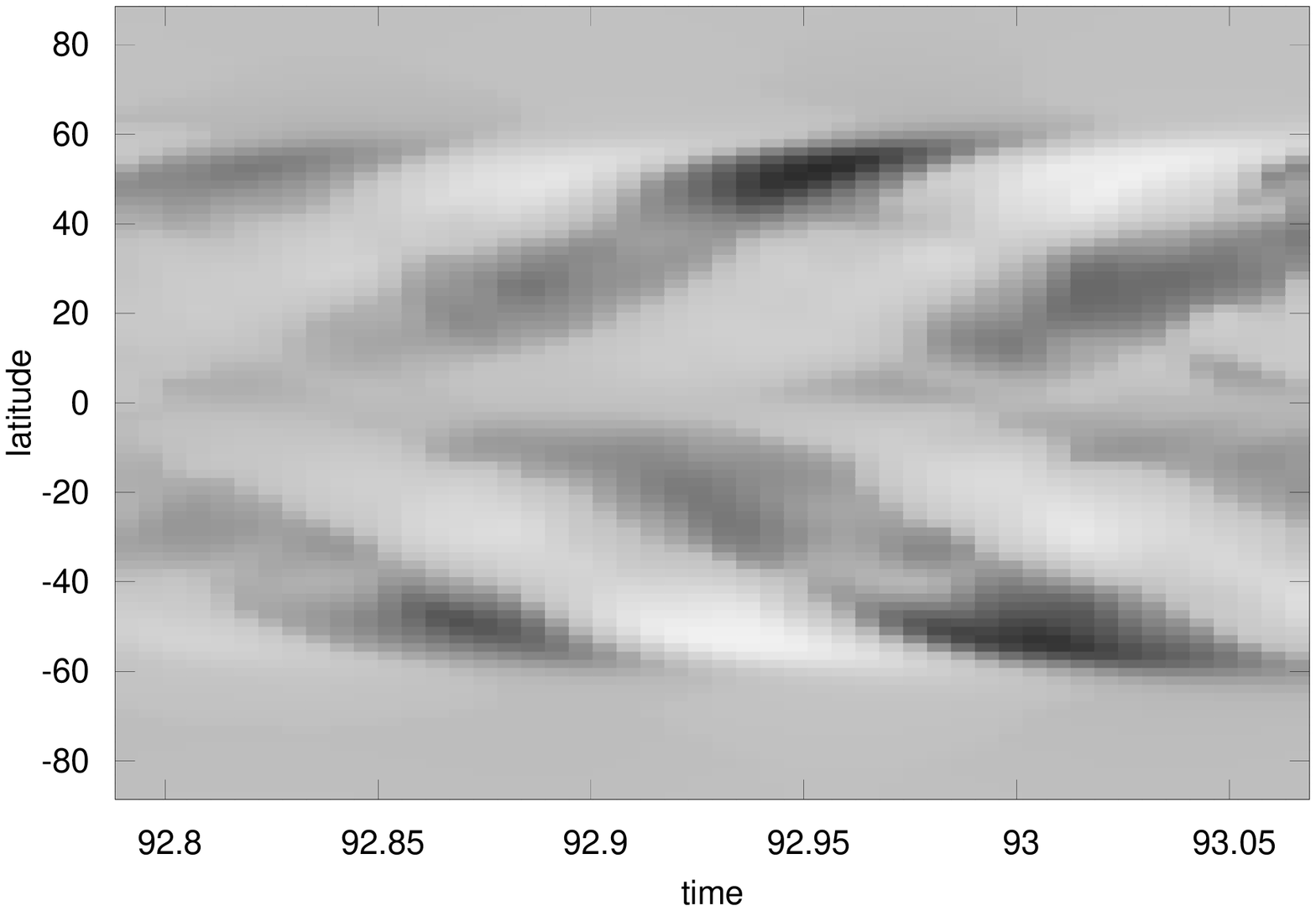,height=\columnwidth,angle=-90,clip=}\\
\epsfig{file=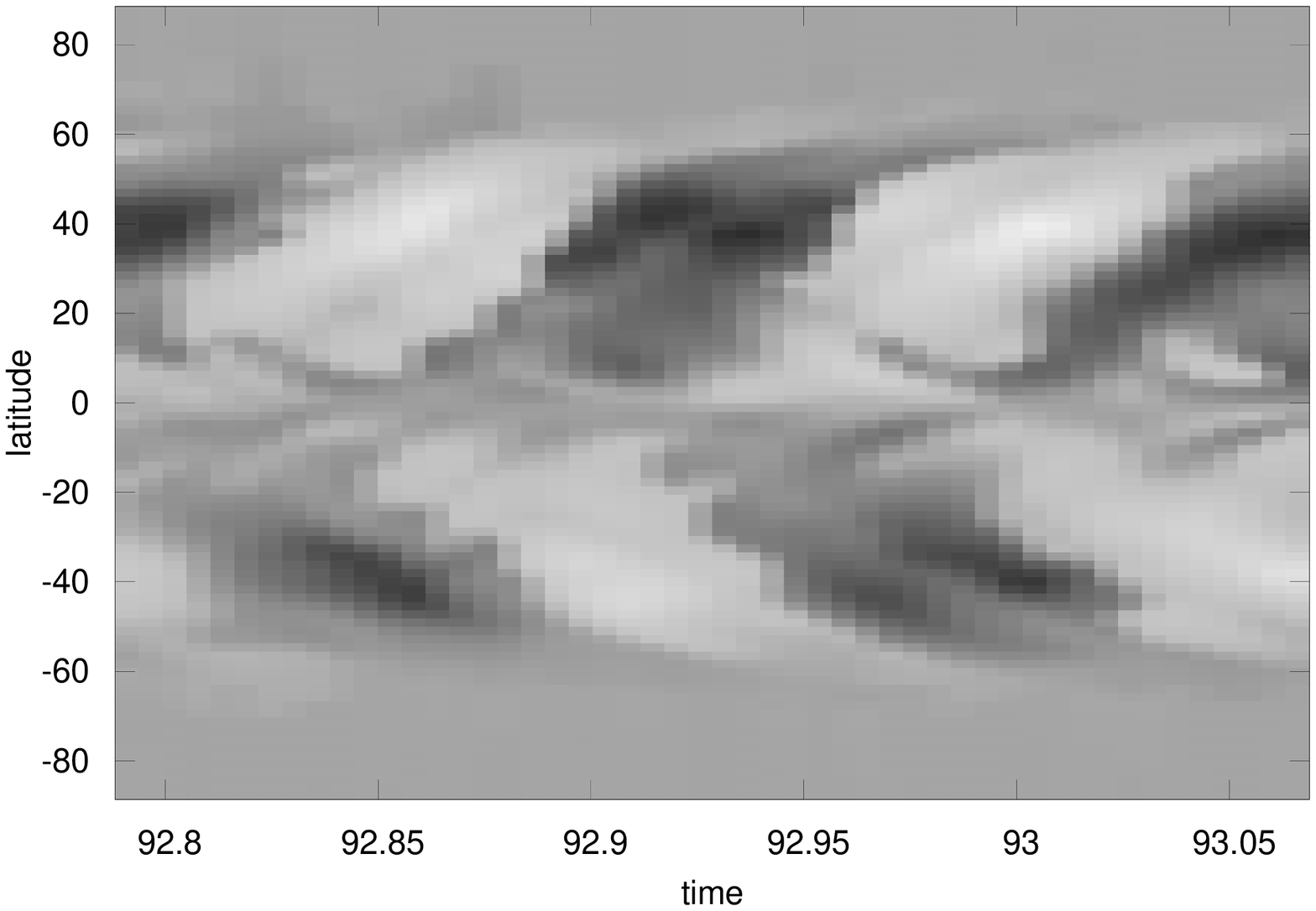,height=\columnwidth,angle=-90,clip=}\\
\end{center}
\caption[]{
"Butterfly" diagrams:
$B_\varphi^{m=0} + |B_\varphi^{m=1}| \sgn(B_\varphi^{m=0})$ (top),
$B_r^{m=0} + |B_r^{m=1}| \sgn(B_r^{m=0})$ (middle), and
$B_\varphi^{m=0} + |B_\varphi^{m=2}| \sgn(B_\varphi^{m=0})$ (bottom)
are plotted as functions of latitude and time in the case $P=1$, $\tau=2000$, $R=140000$, $P_m=3.5$,
 $\beta=1$.
}
\label{fig060}
\end{figure}

It is tempting to {present} visualizations {similar to} the famous
butterfly 
diagrams of solar cycles that have recorded observations of sunspots
for many decades. Since we associate the occurrence of sunspots with
the strength of the horizontal component of the magnetic field in the
sun near the surface, the amplitudes of the maximum values of the
horizontal components could be plotted as function of latitude and
time. But this procedure could not be realized {due to limitations
of} data storage. Instead we have restricted the  attention to the
main contributer to the azimuthal component of the magnetic
field. Thus, in {the top panel of figure \ref{fig060}} the extremal 
azimuthal component of the magnetic field corresponding to the wave
numbers {$m=0$} and {$m=1$}, i.e. $\overline{B_{\varphi}} +
|B^{m=1}_{\varphi}|\sgn(\overline{B_{\varphi}})$, is plotted as
{function} of time and latitude. In contrast to solar butterfly diagrams
the movements to higher latitudes still dominate over the movements
towards lower latitudes. The latter are only visible in the form of
streaks at lower latitudes. 
For comparison the same quantity, but with $m=1$ replaced by $m=2$,
has been plotted in {bottom panel of figure} \ref{fig060}. The
similarity between the two plots indicates that the time and latitude
dependence of the extremal values of $B_{\varphi}$ can approximately
be captured in {this} way. 

In addition to butterfly diagrams based on sunspots also butterfly
diagrams based on observations of the radial component of the magnetic
field are often used. In {the middle panel of figure} \ref{fig060} a theoretical  butterfly
diagram is shown where $\overline{B_r} + |B^{m=1}_r
|\sgn(\overline{B_r})$ has been plotted as function of time and
latitude. A comparison of {the top and the middle panels of figure
\ref{fig060}} indicates a noticeable phase shift between the two
diagrams which agrees roughly with the well known {observation
that the radial component of the solar magnetic field} changes its
sign near the maximum of the solar cycle (Stix, 1976).  

\section{Conclusion}

In this report an attempt is described to explore the extent to which
the operation of the solar dynamo can be understood on the basis of a
minimal, but physically consistent, {convection-driven} dynamo
model. Although the Boussinesq approximation is a highly unrealistic
assumption in the case of the Sun, 
the convection columns are similar to those found in numerical
simulations based on {anelastic} models, see, for instance, Brun
et al.~(2004) and Ghizaru et al.~(2010). Although there is little solar evidence for this type
{of} convection, there are generally believed to exist as "giant cells" in
the deeper region of the solar convection zone. 

The structure of the magnetic field found in our simulations differs
substantially from the commonly assumed structure of the solar
magnetic field dominated by a strong axisymmetric toroidal
component. When the evidence for this traditional view is examined,
however, it is found that observations do not contradict fields
dominated by $m=1$ or $m=2$ components such as those shown in figures
\ref{fig020} and \ref{fig050}. On the contrary, the presence of active
longitudes on the sun indicates at least that components with the
azimuthal wavenumbers $m=1$ and $m=2$ play a significant role.  

{A more detailed exploration of the parameter space of our minimal
model is likely to reveal an even better correspondence with solar
observations.} 

\appendix

\section{Mathematical Formulation of the Numerical Model}

In the numerical model a spherical fluid shell rotating about a fixed
axis described by the unit vector $\vec k$ is considered. It is assumed that a  
static state exists with the temperature distribution
$$ 
%T_S = T_0 - \beta d^2 r^2 /2 + \Delta T \eta r^{-1} (1-\eta)^{-2},
T_S = T_0 + \Delta T \eta r^{-1} (1-\eta)^{-2},
$$
where $r$ denotes the distance from the center of the spherical shell measured in terms of multiples of the shell thickness $d$ and
$\eta$ denotes the ratio of inner to outer radius of the shell. $\Delta T$ is the temperature difference between the two
boundaries.
%$T_S = T_0 - \beta d^2 r^2 /2$. Here $rd$ is the length of
%the position vector with respect to the center of the sphere.
The gravity field is given by 
$
\vec g = - d \gamma \vec r.
$
In addition to  $d$, the
time $d^2 / \nu$,  the temperature $\nu^2 / \gamma \alpha d^4$, and 
the magnetic flux density $\nu ( \mu \varrho )^{1/2} /d$ are used as
scales for the dimensionless description of the problem  where $\nu$ denotes
the kinematic viscosity of the fluid, $\kappa$ its thermal diffusivity,
$\varrho$ its density and $\mu$ is its magnetic permeability.
The equations of motion for the velocity vector $\vec u$, the heat
equation for the deviation 
$\Theta$ from the static temperature distribution, and the equation of
induction for the magnetic flux density $\vec B$ are thus given by 
%\begin{eqnarray}
\begin{gather}
\label{1a}
\partial_t \vec{u} + \vec u \cdot \nabla \vec u + \tau \vec k \times
\vec u = - \nabla \pi +\Theta \vec r + \nabla^2 \vec u + \vec B \cdot
\nabla \vec B, \\
\label{1b}
\nabla \cdot \vec u = 0, \\
\label{1c}
P(\partial_t \Theta + \vec u \cdot \nabla \Theta) = (R\, \eta
r^{-3} (1 - \eta)^{-2}) \vec r \cdot \vec u + \nabla^2 \Theta, \\
\nabla \cdot \vec B = 0, \\
\label{1d}
\nabla^2 \vec B =  P_m(\partial_t \vec B + \vec u \cdot \nabla \vec B
-  \vec B \cdot \nabla \vec u),
\end{gather}
%\end{eqnarray}
where $\partial_t$ denotes the partial derivative with respect to time
$t$ and where all terms in the equation of motion that can be written
as gradients have been combined into $ \nabla \pi$. The Boussinesq
approximation is assumed in that the density $\varrho$ is
regarded as constant except in the gravity term where its temperature
dependence, given by $\alpha \equiv - ( \partial \varrho/\partial T)/\varrho =
const.$, is taken into account. The Rayleigh numbers $R$,
the Coriolis number $\tau$, the Prandtl number $P$ and the magnetic
Prandtl number $P_m$ are defined by 
%3
\begin{equation}
%\R = \frac{\alpha \gamma \beta d^6}{\nu \kappa} , 
R = \frac{\alpha \gamma \Delta T d^4}{\nu \kappa},
\enspace \tau = \frac{2
\Omega d^2}{\nu} , \enspace P = \frac{\nu}{\kappa} , \enspace P_m = \frac{\nu}{\lambda},
\end{equation}
where $\lambda$ is the magnetic diffusivity.  Because the velocity 
field $\vec u$ as well as the magnetic flux density $\vec B$ are
solenoidal vector fields,   the general representation in terms of
poloidal and toroidal components can be used, 
%\begin{eqnarray}
\begin{gather}
\vec u = \nabla \times ( \nabla v \times \vec r) + \nabla w \times 
\vec r \enspace , \\
\vec B = \nabla \times  ( \nabla h \times \vec r) + \nabla g \times 
\vec r \enspace .
\end{gather}
%\end{eqnarray}
Equations for $v$ and $w$ are obtained by multiplication of the
curl$^2$ and of the curl of equation (1) by $\vec r$. Analogously
equations for $h$ and $g$ are obtained through the multiplication of
equation (4) and of its curl by $\vec r$. 

 No-slip boundary conditions are used at the inner boundary and
 stress-free conditions are applied at the outer boundary, while the
 temperature is assumed to be fixed at both boundaries, 
\begin{gather}
%\begin{eqnarray}
\label{ns}
v = \partial_r v = w = \Theta = 0
\quad \mbox{ at } r=r_i \equiv \eta/(1-\eta) 
\\ v = \partial^2_{rr}v = \partial_r (w/r) = \Theta =0
\quad \mbox{ at } r=r_o \equiv 1/(1-\eta).
\end{gather}
%\end{eqnarray}
Only the value $\eta=0.65$ is used in the present paper. An exception
from the stress-free boundary at $r=r_o$ will be assumed for the
axisymmetric part $\overline{w}$ of $w$, 
\begin{equation}
\partial_r (\overline{w}/r) = -\beta \overline{w}/r.
\end{equation}
For the magnetic field, electrically insulating
boundaries are assumed such that the poloidal function $h$ must be 
matched to the function $h^{(e)}$, which describes the  
potential fields 
outside the fluid shell  
%4b
\begin{equation}
%\hspace*{-8mm}
\label{mbc}
g = h-h^{(e)} = \partial_r ( h-h^{(e)})=0 
\mbox{ at } r=r_i  \mbox{ and at }  r=r_o.
\end{equation}
 The numerical integration proceeds with the pseudo-spectral 
method as described by Tilgner (1999)
which is based on an expansion of all dependent variables in
spherical harmonics for the $\theta , \varphi$-dependences, i.e. 
%5
\begin{equation}
v = \sum \limits_{l,m} V_l^m (r,t) P_l^m ( \cos \theta ) \exp(im \varphi)
\end{equation}
and analogous expressions for the other variables, $w, \Theta, h$ and $g$. 
Here $P_l^m$ denotes the associated Legendre functions.
For the $r$-dependence expansions in Chebychev polynomials are used. 
For the computations to be reported in this paper a minimum of
41 collocation points in
the radial direction and spherical harmonics up to the order 128 have been
used.

{
The magnetic energy density components of dynamo solutions are defined as
\begin{gather}
\overline{M}_p = \frac{1}{2} \langle \mid\nabla \times ( \nabla
\overline{h}
\times \vec r )\mid^2 \rangle ,  \quad
 \overline{M}_t = \frac{1}{2} \langle \mid\nabla
\overline g \times \vec r \mid^2 \rangle, \nonumber\\
\check{M}_p = \frac{1}{2} \langle \mid\nabla \times ( \nabla
\check h
\times \vec r )\mid^2 \rangle , \quad
 \check{M}_t = \frac{1}{2} \langle \mid\nabla
\check g \times
\vec r \mid^2 \rangle, \nonumber
\end{gather}
where $\langle\cdot\rangle$ indicates the average over the fluid shell
and $\overline h$ refers to the axisymmetric component of $h$,
while $\check h$ is defined by $\check h = h - \overline h $.
The corresponding kinetic energy densities $\overline{E}_p$,
$\overline{E}_t$, $\check{E}_p$ and $\check{E}_t$
are defined analogously with $v$ and $w$ replacing $h$ and
$g$.}

\section*{{Acknowledgement}}
The research reported in this paper was performed in parts during the
authors' participation in the 2011 program "Dynamo, Dynamical Systems
and Topology" at NORDITA. The research of R.S.~has also been supported
by the UK Royal Society under Research Grant 2010 R2. The research of
F.B.~has also been supported by NASA Grant NNX09AJ85G. The authors are
indebted to Prof.~A.~Kosovichev for his continuing encouragement.

\end{document}